\documentclass[sigconf]{acmart}
\usepackage{graphicx}
\usepackage{subfigure}
\usepackage{multirow}

\usepackage{amssymb}
\AtBeginDocument{%
  \providecommand\BibTeX{{%
    \normalfont B\kern-0.5em{\scshape i\kern-0.25em b}\kern-0.8em\TeX}}}

\setcopyright{acmcopyright}
\copyrightyear{20XX}
\acmYear{20XX}
\acmDOI{10.1145/XX}

\begin{document}

\title{A parallel-network continuous quantitative trading model with GARCH and PPO}

\author{Zhishun Wang}
\affiliation{%
	\institution{University of Electronic Science and Technology of China}
	\country{}}

\author{Wei Lu}
\affiliation{%
	\institution{University of Electronic Science and Technology of China}
	\country{}}

\author{Kaixin Zhang}
\affiliation{%
	\institution{Harbin Institute of Technology}
	\country{}}
\author{Tianhao Li}
\authornote{corresponding author}
\email{tinho1995@sina.com}
\affiliation{%
	\institution{University of Electronic Science and Technology of China}
	\country{}}
\author{Zixi Zhao}
\affiliation{%
	\institution{Harbin Institute of Technology}
	\country{}}
%\author{Zhishun Wang}
%\affiliation{%
%  \institution{University of Electronic Science and Technology of China}
%  \country{}}
%
%\author{Wei Lu}
%\affiliation{%
%  \institution{University of Electronic Science and Technology of China}
%  \country{}}
%
%\author{Kaixin Zhang}
%\affiliation{%
%  \institution{Harbin Institute of Technology}
%  \country{}}
%\author{Tianhao Li}
%\authornote{corresponding author}
%\email{tinho1995@sina.com}
%\affiliation{%
%  \institution{University of Electronic Science and Technology of China}
%  \country{}}
%\author{Zixi Zhao}
%\affiliation{%
%	\institution{Harbin Institute of Technology}
%	\country{}}

%%\renewcommand{\shortauthors}{Trovato and Tobin, et al.}

\begin{abstract}
It is a difficult task for both professional investors and individual traders continuously making profit in stock market. With the development of computer science and deep reinforcement learning, Buy\&Hold (B\&H) has been oversteped by many artificial intelligence trading algorithms. However, the information and process are not enough, which limit the performance of reinforcement learning algorithms. Thus, we propose a parallel-network continuous quantitative trading model with GARCH and PPO to enrich the basical deep reinforcement learning model, where the deep learning parallel network layers deal with 3 different frequencies data (including GARCH information) and proximal policy optimization (PPO) algorithm interacts actions and rewards with stock trading environment. Experiments in 5 stocks from Chinese stock market show our method achieves more extra profit comparing with basical reinforcement learning methods and bench models.
\end{abstract}




\maketitle

\section{Introduction}
Stock market plays an important role in finance. Investors all expect to get higher returns in stock trading, but stock trading is affected by many factors, such as information, volatility, leverage, etc. Meanwhile, stock market data is complex, noisy, nonlinear, and non-stationary \cite{Hadi2021}. Traditional quantitative trading methods have faced more challenges while dealing with such difficulties. With the development of machine learning methods in recent years, the combination of machine learning and quantitative trading has shown great potential for returns. Due to the ability of machine learning algorithms that capture richer indicators and information, excess returns can be obtained.

\par Reinforcement learning (RL), as one of the popular machine learning methods, is widely used in stock trading decision-making. RL employs an agent to interact with the environment. The goal of RL is maximizing its cumulative return, which is consistent with the scene of stock trading. Existing studies acquired appealing achievements in stock trading applications compared with traditional machine learning methods \cite{Salvatore2021}. However, there are several problems listed need to be improved according to the newest researches.

\par Firstly, Trading with a single stock history or indicator is a common work in existing works \cite{Luo2019}, such as closing price, moving average indicator, and so on, which can not provide more plentiful information for trading algorithms needed. Indicator selection is crucial for stock trading algorithms, which determine the ultimate algorithms' performance, especially in deep learning (DL) and RL. So how to make use of the effective trading information is the bottleneck of the algorithm.

\par Secondly, in the existing stock trading papers with RL, their output actions are discretized into trading signals, usually including Buy \& Hold \& Sell \cite{Jagdish2021,Hirchoua2021}. Such signals can only decide the direction of trading, not the number of trading shares. Therefore, these algorithms are greatly limited in practical trading.

\par To tackle these challenges, we propose a \underline{m}ulti-frequency \underline{c}ontinuous-share quantitative \underline{t}rading algorithm with \underline{G}ARCH (MCTG) based on proximal policy optimization algorithm (PPO)\cite{Sutton1998}, which is a policy gradient method of reinforcement learning. Firstly, the state space of our MCTG model contains the information of multi-frequency data processed by parallel network layers. The multi-frequency data containing 3 different periods (5 min, 1 day, and 1 week) can provide more abundant information for DL. And the parallel network layer is designed o capture more valid information and guide more valuable trading decisions. Secondly, we use volatility as a measure of risk and consider the volatility prediction Model GARCH which is widely used in econometrics. The volatility data predicted by GARCH is used to supplement the daily frequency input data, which is helpful for the agent to recognize risk, reduce transaction taxes and obtain higher returns. Finally, as for our action space of RL, we build a continuous trading decision algorithm, which can continuously buy and sell stock shares in the range number of -1 to +1, with negative numbers indicating selling stock and positive indicating to buy. And the flexible trading according to the proportion of output actions is conducive to the ability of multi-frequency data combined with GARCH model to measure risk, reduce transaction taxes and obtain higher returns. Our experimental results show that the proposed model can significantly outperform the bench model on five stocks.

\begin{figure}[htbp]
	\centering
	\includegraphics[width=0.54\textwidth,trim=70 80 40 105,clip]{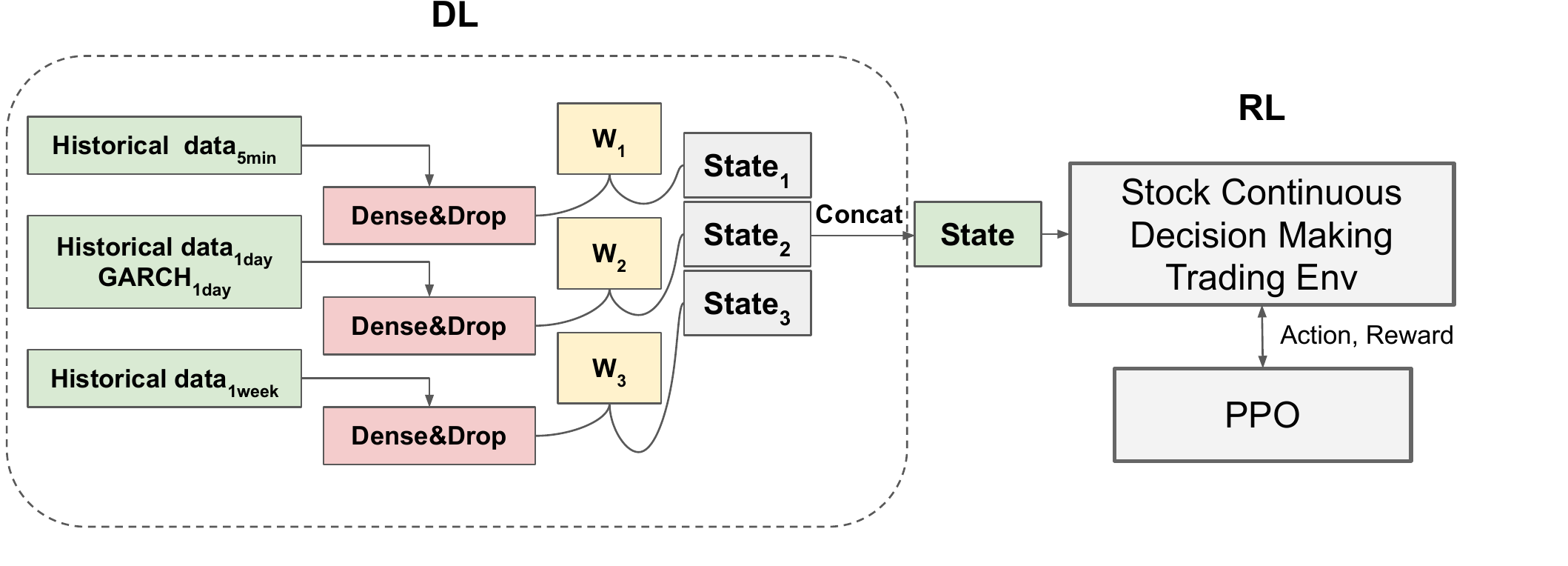}
	\caption{The framework of MCTG model with DL data processing and RL continuous decision making.}
\end{figure}

\section{Prolem Statement}
We formulate the stock trading process into a Markov Decision Process (MDP) in order to match with DRL algorithms. A five-tuple array include <$s, p, a, r$, $\gamma$>, which are defined as state  space $s$, transition function  $p$, action space $a$, reward function $r$ and discount factor $\gamma$. The state transition probability is objectively given and is not affected by actions of the agent. Considering the dynamic and uncertain characteristics of the stock market, We set up the stock trading model as an MDP as follows:

\begin{itemize}
	\item {State $s$}: State includes a series of features of the stock's current state. We use the stock's information include six different frequencies of opening, closing, high, low, volume, amount and volatility, to build the parallel network layer to form our state.
	\item {Action $a$}: Our agent can take an action set from -1 to 1, which means sell \& buy (0 indicate hold) with a continuous trading decision. For example, the output number of 0.3 means that 30\% of the current cash is used to buy stocks, and - 0.5 means that 50\% of the existing stocks are sold.
	\item {Reward $r(s,a)$}: An agent gets reward according to its trading strategy.
\end{itemize}

\section{Mehtods}
In this section, We specifically introduce the design of our parallel network structure, RL algorithm, and the application of the volatility prediction model.
\subsection{Multi-frenquency parallel architecture}
We make multi-frequency data of stock history trading data (5 min, 1 day with GARCH and 1 week) as the input data of our model. Each of them will be processed by proprietary DNN layers, dropout layer, and all of the layers are called parallel network \cite{Guo2019}. After processed by DL algorithms, there will be three output matrixes with same shapes which include 16 elements, and then they were multiplied with three weight matrices with the same shape are updated gradually with the training of deep reinforcement learning ($W_1,W_2,W_3$). After that, all these three matrixes ($State_1,State_2,State_3$) will be concatenated into a matrix which shape includes 48 elements. So the equation of state $S_t$ computes as follow:
\begin{equation}
\begin{aligned}	
	State_i &= W_i \times S_{i,t},\\
	S_t = Concat &[State_1,State_2,State_3],
\end{aligned}
\end{equation}
where $S_{i,t} (i=1,2,3)$ represents the processed data at different frequencies at times $t$, and $W_i$ represents their weights. More details are shown in Figure 1.
\subsection{Proximal Policy Optimization Algorithm }
The reinforcement learning algorithm can be divided into two types, the value-based method and the policy gradient (PG) method. The former is based on the Q value of a state-action pair $(s, a)$, which is not suitable for our problem. The latter optimizes the policy directly by using gradient ascending to optimize the objective function, this algorithm is more suitable for continuous trading decisions. The formula of PG is shown as below:
\begin{equation}
	\nabla_\theta J(\pi_\theta)=\mathop{\mathbb{E}}_{\tau\sim\pi_\theta}\left[ \sum_{t=0}^{T} \nabla_\theta log\pi_\theta(a_t|s_t)A^{\pi_\theta}(s_t,a_t)       \right],
\end{equation}
where $\pi _{\theta}$ denotes a policy with parameters $\theta$, $J(\pi_{\theta})$ denotes the expected finite-horizon undiscounted return of the policy, $\tau$ is a trajectory and $A^{\pi_{\theta}}$ is the advantage function for the current policy. In PPO algorithm\cite{John2017}, it defines a probability ratio:
\begin{equation}
	\rho(\theta)= \frac{\pi_\theta(a|s)}{\pi_{\theta_k}(a|s)},
\end{equation}
and try to optimize the following objective function:
\begin{equation}
L(s,a,\theta_k,\theta) = \min\left(
\rho(\theta) A^{\pi_{\theta_k}}(s,a),
\text{clip}\left(\rho(\theta), 1 - \epsilon, 1+\epsilon \right) A^{\pi_{\theta_k}}(s,a)
\right),
\end{equation}
where $\epsilon$ is a very small hyperparameter which roughly says how far away the new policy is allowed to go from the old.

\subsection{Generalized Autoregressive Conditional Heteroskedasticity}
The Generalized Autoregressive Conditional Heteroskedasticity (GARCH) model describes the variance of the current error terms as a function of the previous periodic error terms\cite{HU2020}. The specification for the GARCH$(p, q)$ is defined as:

\begin{equation}
\begin{aligned}	
R_t &= \mu+ a_t ,\\
a_t &= \varepsilon_t\sigma_t ,\\
\sigma_t^2 &= \alpha_0+\sum_{i=1}^{p}\alpha_i a_{t-i}^2+\sum_{j=1}^{q}\beta_j a_{t-j}^2,
\end{aligned}
\end{equation}
in existing studies, the rolling prediction of GARCH and the stock return rate has a time correlation \cite{Helmut2017}. In this paper, we choose the GARCH (1,1) model to make a rolling prediction of volatility, and then add the volatility into our day level data.
\section{Experiments}
\subsection{Experiments setting}
\textbf{Dataset}. We collect five stocks from different sectors in Chinese stock market (SZ.002230, SZ.000333, SH.603288, SH.600030, SH.600276, SH and SZ represent the Shanghai and Shenzhen stock exchange), and the range of data is 2011 to 2020 with different frequency (5 min, 1 day with GARCH and 1 week). We divide data into training set and test set. The data from 2011 to 2018 is used as the training set, and the rest is test set.\\
\textbf{State space}. Our state space contains history multi-frequency trading information and GARCH volatility items. The multi-frequency information includes 5 min, 1 day with GARCH and 1 week for short, medium and long term information which can provide more abundant information than single daily frequency data. As the trading time is 4 hours a day, including 48 sets of 5 min short-term information, the medium-term information consists of 30 daily frequency information and rolling GARCH fluctuation items, and the long-term information consists of 30 weekly frequency information. The multi-frequency information is represented by the three blue parts on the left in Figure 1.\\
\textbf{Action space}. We employ our agent firstly to output a continuous value $a_t$ from -1 to 1, which specifies both the trading directions (buying and selling determined by signs plus and minus) and the number of stock shares. The flexible trading signals are conducive to the ability of the output of state space to measure risk, reduce transaction taxes and obtain higher returns. The final output is an action that determines buy or sell on the current state computed by (6). We consider the design of minimum transaction with 100 shares and the transaction fees charged in Chinese market. To ensure the validity of the trade, we use the next day's opening prices of the stock as the buying and selling point and the formula for final action as follows:

\begin{equation}
	 A_t=\left\{
	\begin{aligned}
	&\left[\frac{cash_t \times a_t}{opening_t \times (1 + tax)}\right]&\text{if}\quad a_t \ge 0 \\
	&\left[\frac{hold_t \times a_t}{opening_t}\right]&\text{if}\quad a_t < 0,
	\end{aligned}
\right.
\end{equation}
where $[\cdot]$ represents an integral function. Cash, tax and hold represent the amount of cash, the  transaction tax and the stock shares holding. \\
\textbf{Reward function}. To get more extra profit rate comparing with the B\&H baseline, we consider making our reward function as follows:
\begin{equation}
	Reward_t=\frac{hold_t - hold_{t-1}}{hold_{t-1}}-\frac{opening_t-opening_{t-1}}{opening_{t-1}},
\end{equation}
the first term of equation (7) represents the return rate of assets held, and the second term represents the return rate of stock price. The design of reward sufficiently enables the agent to learn to control its position through continuous decisions, which better fits the trading logic of real investors.\\
\textbf{Hyperparameters}. The proposed model adopts Stable-Baselines and TensorFlow framework. In the PPO algorithm, we select the learning rate, batch size, discount rate and minibatch are 0.00025, 1024, 0.99, 4. In the parallel network layer, three DNN models dealing with different frequencies are all adopted four-layer neural networks, with hidden layers of 32 and 16 units. We also add a dropout layer with rate of 0.25.
\subsection{Results}
We compare 4 different DL methods based on PPO including DNN (Deep neural network), DNN-GARCH (Deep neural network with GARCH), MCT (Multi-frequency continuous-share trading model) and MCTG (Multi-frequency continuous-share trading model with GARCH). Our method is compared with B\&H as a baseline in each model.\\
\textbf{Training performance analysis}. The training process is shown in Figure 2. The final result of the episode reward has been magnified 100 times for a better visual presentation. During the beginning of training, the reward of the four models shows an increasing trend. As we continue to train, MCT and MCTG were significantly better than DNN and DNN-GARCH in episode rewards on the four groups of experiments. Finally, MCTG achieves the best training effect, which indicate that the GARCH model provides more useful information during the training process.\\
\textbf{Testing performance analysis}. After trading 2 years from 2019 to 2020, we get the performance of the algorithm which is shown in Table 1 and Figure 3. In Table 1, it includes the profit rate (PR) and tax rate (TR) performance of five stocks under different models. Profit rate (PR) represents the annual return rate. Tax rate (TR) represents the annual tax rate, which is used to describe the transaction frequency. There is no TR since the B\&H strategy does not trade.
\begin{figure}[tb]
	\centering
	\includegraphics[width=0.5\textwidth]{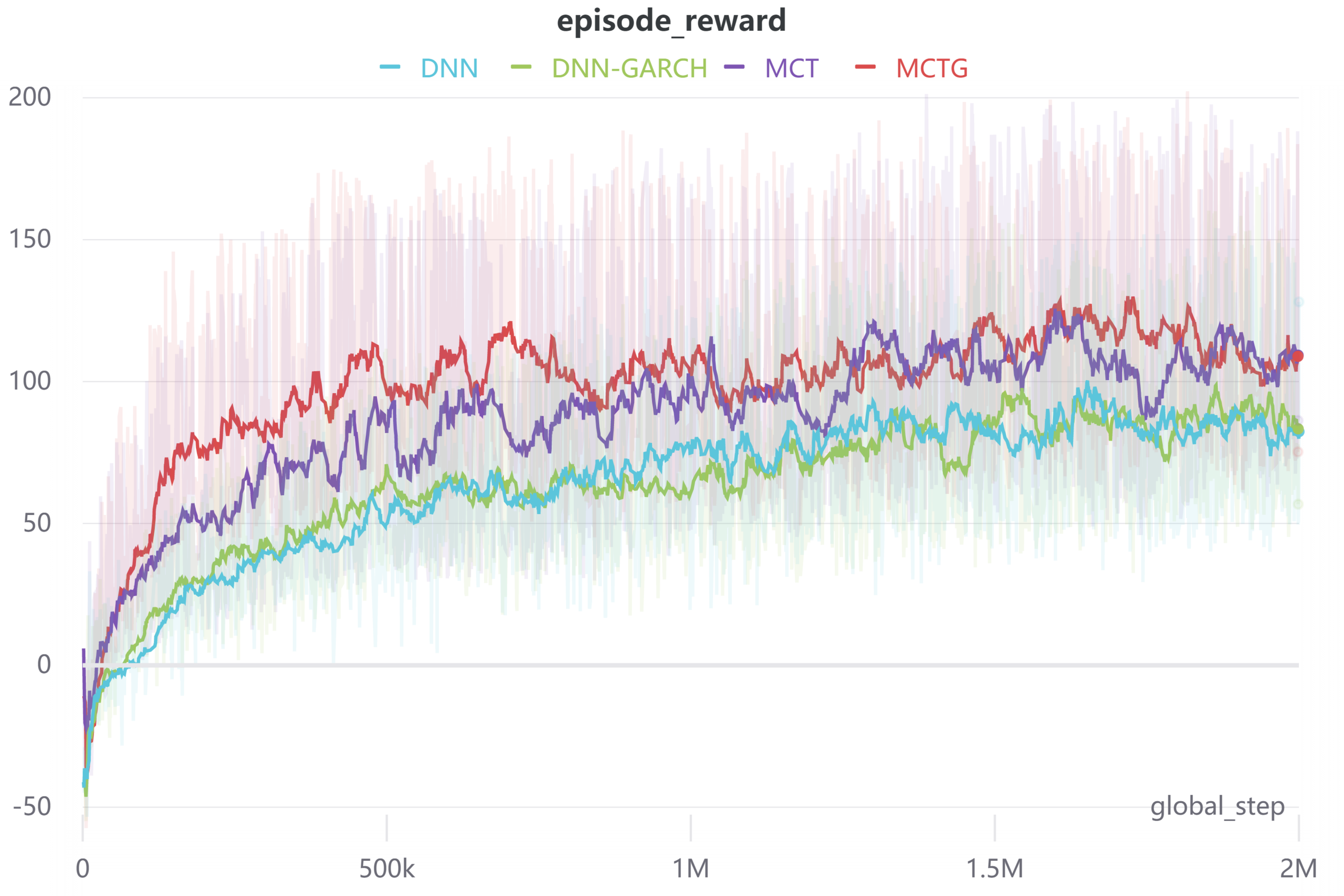}
	\caption{Reward over 4 models in 2 million training steps.}
\end{figure}
\begin{table*}[t]
	\caption{Experiment results of different models}
	\begin{tabular}{cccccccccc}
		\hline
		& B\&H                         & \multicolumn{2}{c}{DNN}               & \multicolumn{2}{c}{DNN-GARCH} & \multicolumn{2}{c}{MCT}                 & \multicolumn{2}{c}{MCTG} \\ \cline{2-10}
		\multirow{-2}{*}{Model}        & PR                           & PR                           & TR                          & PR                                 & TR            & PR                            & TR                           & PR                               & TR         \\ \hline
		SZ002230 & 13.83\% & 19.85\% & 5.46\%                      & 14.45\%       & 2.31\%        & \textbf{65.24}\%  & 13.66\%                      & 62.44\%     & 12.25\%    \\
		SZ000333 & 64.02\% & 65.77\% & 7.72\%                      & 64.17\%       & 10.40\%       & 98.70\%  & 11.40\%                      & \textbf{106.04}\%    & 11.91\%    \\
		SH603288 & 88.42\% & 84.50\% & 8.82\%                      & 98.49\%       & 6.79\%        & 115.02\% & 19.27\%                      & \textbf{144.20}\%    & 18.02\%    \\
		SH600030 & 34.58\% & 45.46\% & 11.45\%                     & 54.50\%                            & 7.62\%        & 75.35\%  & 16.46\%                      & \textbf{84.24}\%     & 16.55\%    \\
		SH600276 & 69.24\% & 79.18\%                      & 7.98\% & 88.26\%                            & 9.58\%        & 99.82\%                       & 17.87\% & \textbf{101.67}\%                         & 15.04\%    \\ \hline
		mean     & 54.02\%                      & 58.95\% & 8.29\%                      & 63.97\%       & 7.34\%        & 90.83\%  & 15.73\%                      & \textbf{99.72}\%     & 14.75\%    \\ \hline
	\end{tabular}
\end{table*}
\begin{figure*}[h]
	\centering
	\subfigure{
		\includegraphics[width=5.7cm,trim=20 0 40 20,clip]{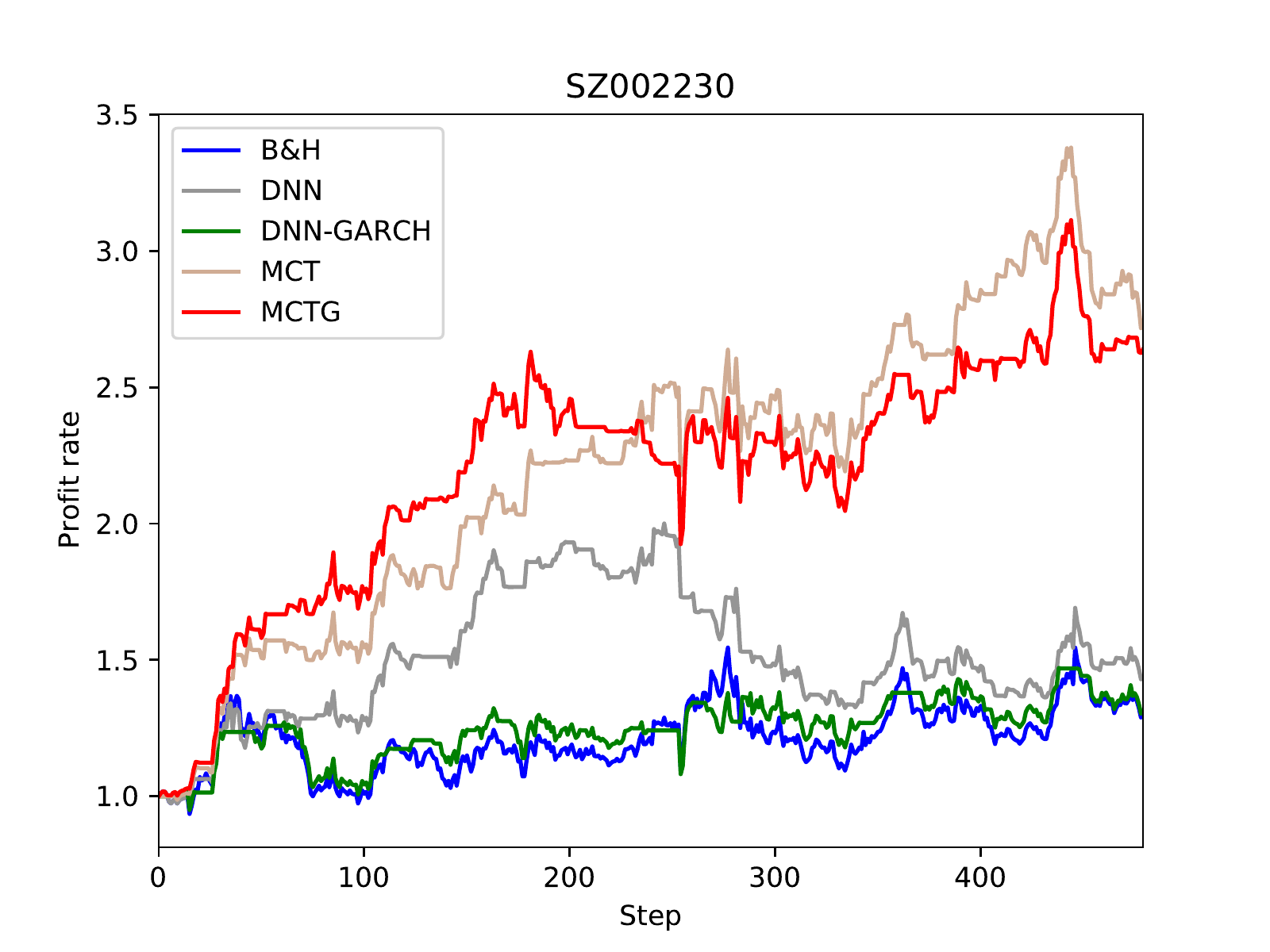}
		%\caption{fig1}
	}
	\subfigure{
		\includegraphics[width=5.7cm,trim=20 0 40 20,clip]{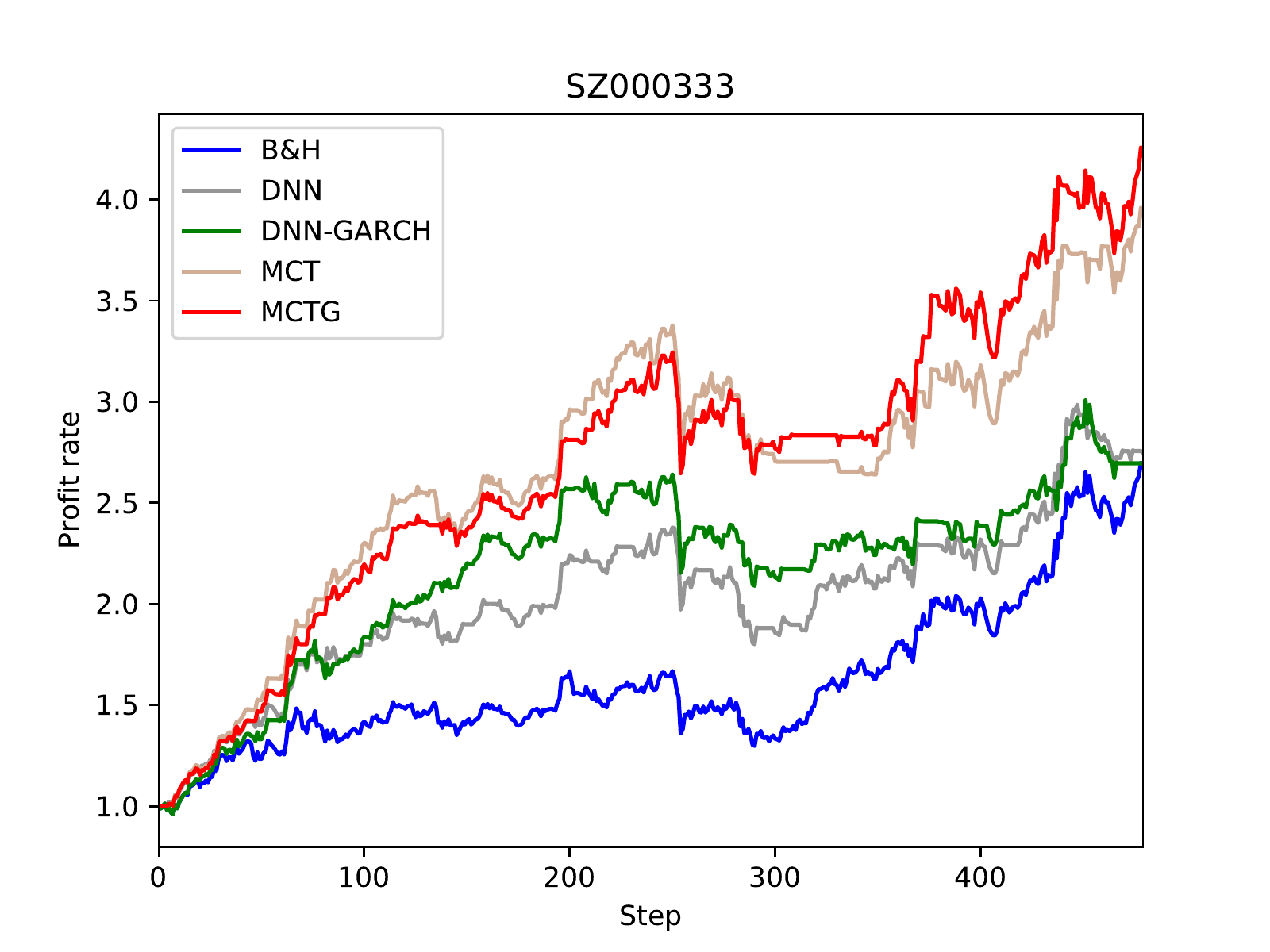}
	}
	\subfigure{
		\includegraphics[width=5.7cm,trim=20 0 40 20,clip]{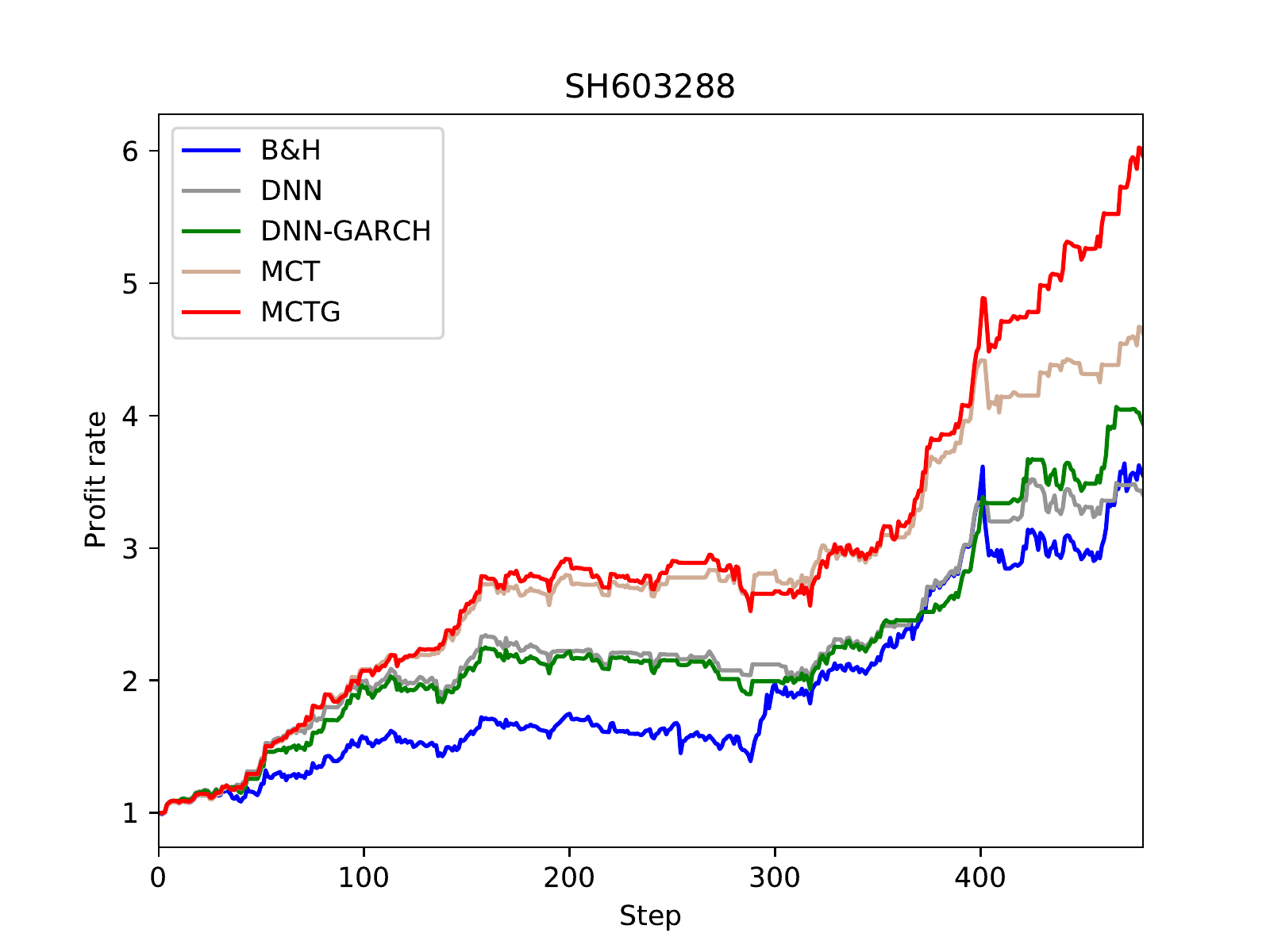}
	}
	
	\subfigure{
		\includegraphics[width=5.7cm,trim=20 0 40 20,clip]{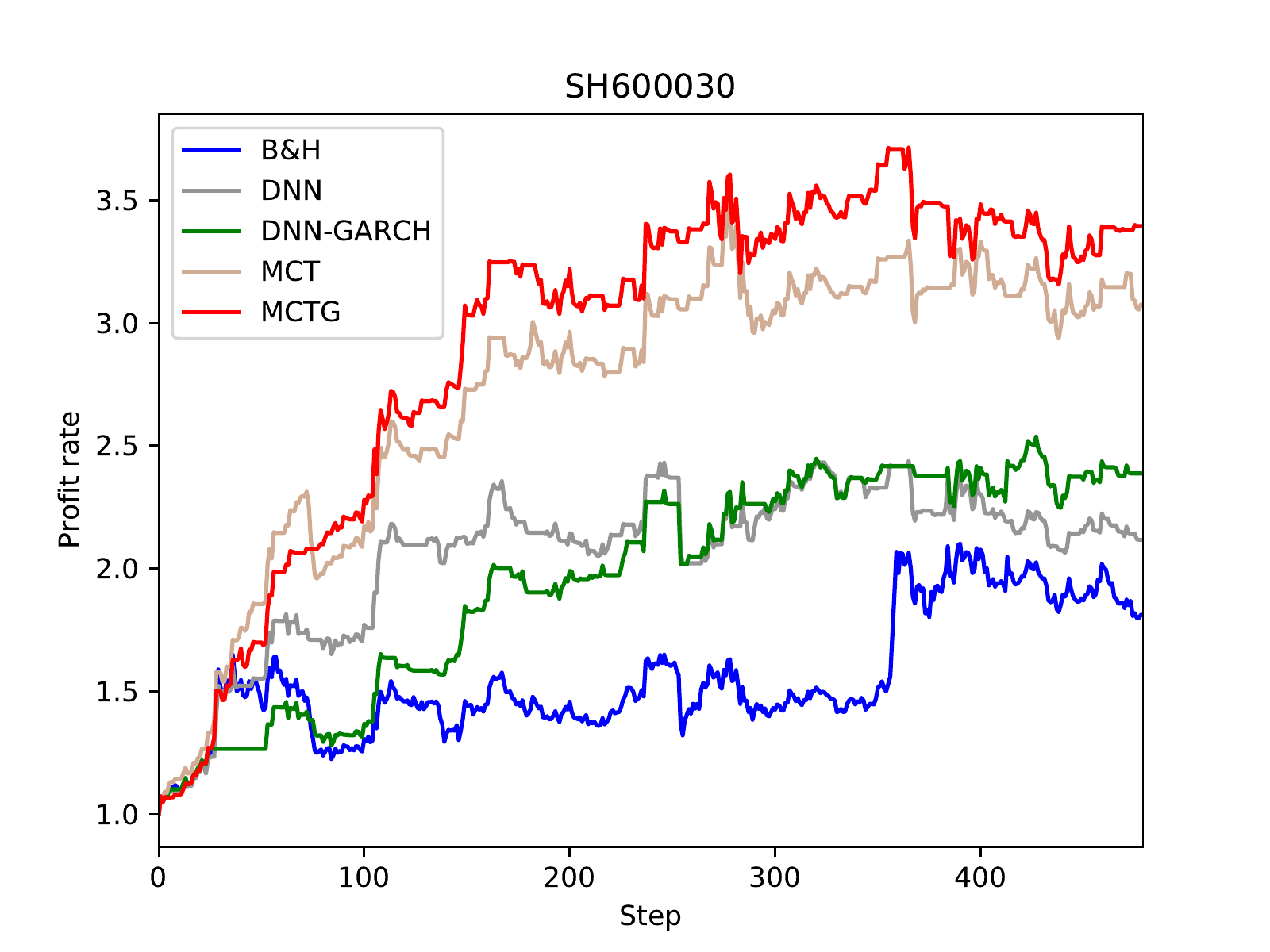}
	}
	\subfigure{
		\includegraphics[width=5.7cm,trim=20 0 40 20,clip]{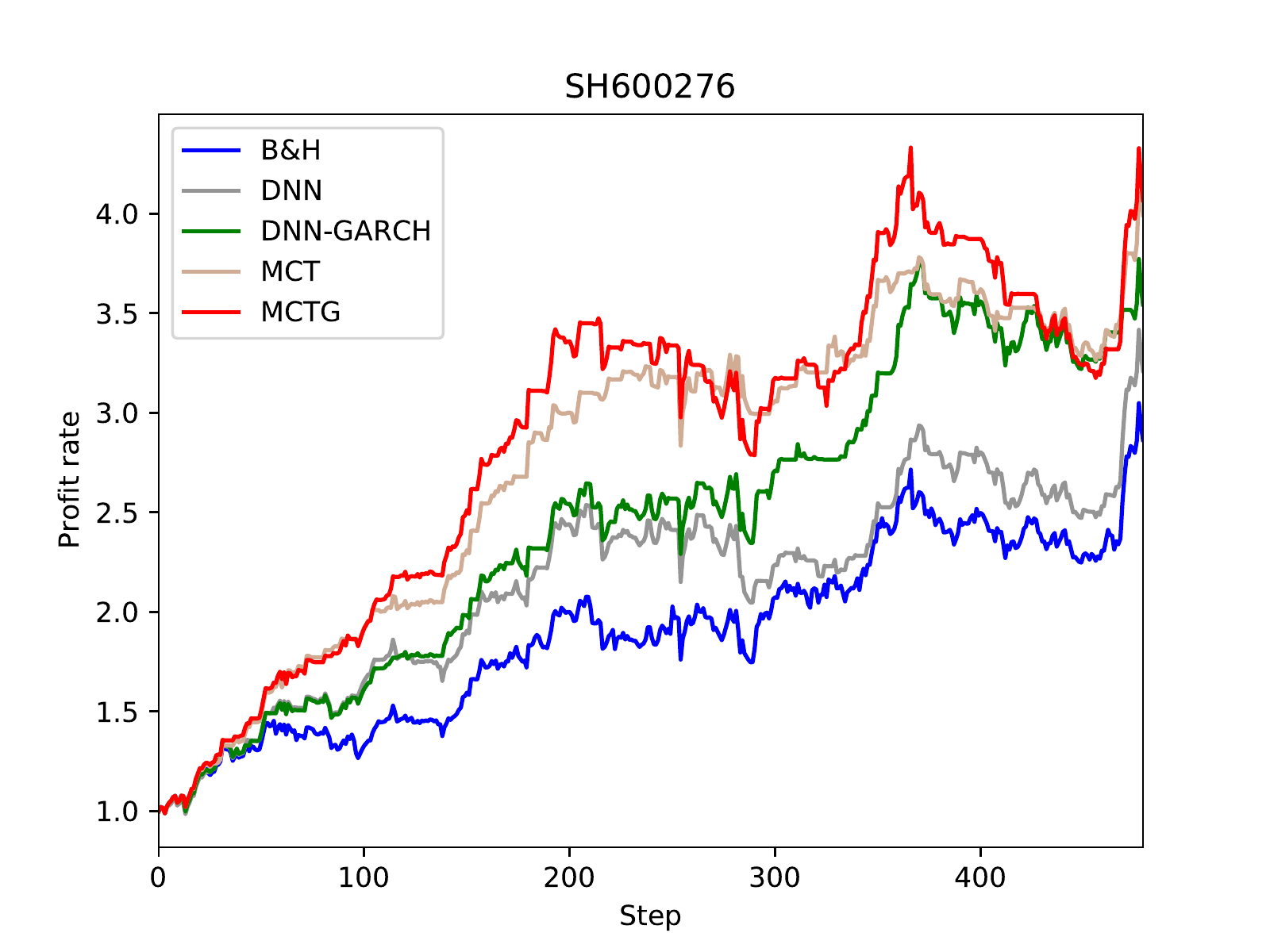}
	}
	\caption{Profit rate over 5 stocks in testing during 2 years from 2019 to 2020.}
\end{figure*}
About the testing performance of five stocks, the DNN model achieves a better PR than B\&H on four stocks and average return. After the addition of GARCH item in DNN model, the average return rate is up to 63.97\%, and the transaction frequency is reduced by nearly 1\% at the same time, which shows that GARCH provides more abundant fluctuation information for agents, thus reducing the transaction TR. PR of the MCT model using multi-frequency information is significantly higher than the DNN models because multi-frequency data and parallel network layer contribute to providing and analysing more effective information for the agent, but TR is higher which is because of the large trading stock shares and more actions to get more trading revenue. The return rate of the MCTG model with GARCH item has reached a higher level nearly 9\% than MCT and the reason is that GARCH model provides more favorable information to help the algorithm to measure risk, reduce transaction taxes and obtain higher returns. Meanwhile, TR has been reduced by nearly 1\%.
\par Figure 3 shows the performance of five stocks during two years test. It can be clearly seen that among the top four stocks (except SH600276), MCTG and MCT have successively achieved higher returns than DNN and B\&H with stable performance, as well as DNN and DNN-GARCH also achieve better performance than B\&H. From the profit rate performance in Table 1, we can see that the return rate of the MCT model on the first stock (SZ002230) is higher than MCTG, and we can find the reason in Figure 3 that the return of the second year does not catch up with MCT, leading to the final annualized return not beating MCT. In the remaining four stocks, MCTG model eventually achieves higher returns than MCT. Especially in the third stock (SH603288), the MCTG model's profit rate is nearly twice as much as DNN and B\&H. All results show the MCTG model has the most powerful performance.

\section{Conclusion}
In this work, we propose a quantitive trading model named the multi-frequency continuous-share trading model with GARCH (MCTG) which contains three parallel network layers and RL. The three parallel network layers process input data with different frequencies and add GARCH item to the daily data to construct our model. Our action space of RL to make continious trading decisions of stock shares is more suitable for practical application. Experiments with two years of five stocks in the Chinese market show that MCTG can effectively process and analyze multi-frequency information, measure risks, reduce transaction taxes, and obtain more benefits. MCTG also outperforms most of other models including the B\&H baseline in terms of PR. We also compared the performance of TR under the same structure models where GARCH contributes to reducing PR. MCTG significantly increases PR and reduces TR. The above results show that our model can better fit the real continuous transactions than other algorithms.
%\begin{acks}
%This work was supported by the National Key Research and
%Development Program of MOST of China under Grant 2018AAA0101003.
%\end{acks}

\newpage
\bibliographystyle{unsrt}
\bibliography{sample-base}

\end{document}